\documentclass[aps,prb,reprint,showpacs,floatfix,twocolumn]{revtex4-1}
\usepackage{amsmath,graphicx,epsfig,color,verbatim,amsfonts,mathtools,amssymb}
\usepackage{tikz}
\usepackage{array}
\usepackage{standalone}
\usepackage{longtable}
\usepackage{bm}
\usetikzlibrary{positioning}

\let\oldhat\hat
\renewcommand{\vec}[1]{\mathbf{#1}}
\renewcommand{\hat}[1]{\oldhat{\mathbf{#1}}}

\begin{document}
%----------------------------------------------------------------------------------------
%	TITLE SECTION
%----------------------------------------------------------------------------------------
\title{New class of planar ferroelectric Mott insulators via first principles design}
\author{Chanul Kim}
\affiliation{Department of Applied Physics and Applied Mathematics, Columbia University, New York, New York 10027, USA}
\author{Hyowon Park}
\affiliation{Department of Physics, University of Illinois at Chicago, Chicago, Illinois, 60607, USA}
\author{Chris A. Marianetti}
\affiliation{Department of Applied Physics and Applied Mathematics, Columbia University, New York, New York 10027, USA}
\email{ck2608@columbia.edu}

%----------------------------------------------------------------------------------------
%	ABSTRACT
%----------------------------------------------------------------------------------------

\begin{abstract}
\noindent
Inorganic perovskite oxide ferroelectrics have recently generated substantial interest for photovoltaic applications; however, existing materials suffer from excessive electronic band gaps and insufficient electric polarization. The recent resurgence of hybrid perovskite ferroelectrics addresses the aforementioned deficiencies, but they are highly unstable against environmental effects and an inorganic resolution may still be optimal. Here we use first-principles calculations to design a low band gap, planar ferroelectric by leveraging the complexity of layered double perovskite oxides AA$^\prime$BB$^\prime$O$_6$.  We exploit A$\ne$A$^\prime$ size mismatch, a nominally empty $d$-shell on the B-site, and A cations bearing lone-pair electrons to achieve a large ferroelectric polarization. Additionally, B$^\prime\ne$B is chosen to achieve full charge transfer and a Mott susceptible filling on the B$^\prime$-site, yielding a low band gap. These principles are illustrated in BaBiCuVO$_6$, BaBiNiVO$_6$, BaLaCuVO$_6$, and PbLaCuVO$_6$, which could be realized in layer-by-layer growth. This new class of materials could lead to stable, high-efficiency photovoltaic.  
\end{abstract}

\date{\today}
%\pacs{62.20.M-, 62.23.Kn, 63.22.-m, 64.70.Nd}
\maketitle

%----------------------------------------------------------------------------------------
%	ARTICLE CONTENTS
%----------------------------------------------------------------------------------------

%\begin{multicols}{2} % Two-column layout throughout the main article text
\section{Introduction}
Ferroelectric oxides have emerged as a candidate for applications in photovoltaic devices, which separate photo-excited electron-hole pairs via the internal electric field of the ferroelectric\cite{Fridkin2012}. Recent progress in domain engineering of ferroelectric oxides has demonstrated above-bandgap open circuit voltages\cite{Yao2005, Yang2010}, in contrast to classic p-n junction photovoltaic\cite{Shockley1961}. Furthermore, ferroelectric oxides are tolerant to various environments and can be fabricated using cost efficient methods\cite{Grinberg2013,Cao2012}. However, their power conversion efficiencies (PCE) have been quite limited until recently, despite decades of exploration in this class of materials\cite{Yuan2014, Grinberg2013, Zhang2013, Nechache2014}. The limitation of many existing ferroelectric oxides is a large electronic band gap (ie. $\ge$ {\it 3}$eV$)\cite{Yuan2014}, which is typical of $d_0$ systems where the gap is between the oxygen $2p$ and the nominally empty transition-metal $d$ states. Band gaps of this magnitude are problematic, as they cannot capture a significant component of the visible spectrum\cite{Shockley1961}. An additional limitation is the extremely low photo-current observed in most ferroelectric oxides\cite{Yuan2014}, and this problem was long hindered by a lack of a detailed mechanistic understanding. However, the application of shift current theory has provided significant insight into this phenomenon, and it is apparent that  strong polarization and covalency are necessary ingredients for enhancing the photocurrent\cite{Young2012,Young2012a}.

In order to circumvent the aforementioned limitations, one must simultaneously engineer a low band gap, a large ferroelectric polarization, and strong hybridization.  Indeed, designing a  lower band gap in ferroelectric oxides has drawn a large degree of attention\cite{Choi2009,Choi2012,Basu2008,Grinberg2013,Zhang2013}.  One possible route to smaller band gaps is to utilize Mott or charge-transfer insulators, as in the well-known case of BiFeO$_3$\cite{Choi2009} where an optical band gap of {\it 2.7}$eV$ is realized.  While the gap is still excessive in this case, this is a promising direction that can be furthered in the context of alloys.  In particular, one can explore the double perovskite A$_{2}$BB$^\prime$O$_6$ in order to combine both aspects of band insulators (ie. nominal $d_0$ valence) with Mott insulators (ie. nominal $d_n$ valence) to achieve a high degree of tunability of the band gap.  A large ferroelectric polarization in the presence of strong hybridization can be achieved via the second order Jahn-Teller effect, as is often realized when A-site cations contain lone-pair electrons and or when B-site ions are in a nominal $d_0$ valence. Furthermore, the magnitude of the ferroelectric polarization can be further enhanced by employing size mismatch on the A/A$^\prime$-sites, particularly when one of the cations is lone-pair active\cite{Singh2008,Takagi2010,Takagi2010a,Shishidou2004}.  In order to maximally exploit all aforementioned mechanisms, one can consider the more general double perovskite AA$^\prime$BB$^\prime$O$_6$\cite{Young2015, Rondinelli2012, Mulder2013}. Such compounds can be realized in experiment via chemical synthesis\cite{Zubko2011,Fukushima2011,Knapp2006,Grinberg2013} and potentially via layer-by-layer growth such as Molecular Beam Epitaxy (MBE) and Pulsed Laser Deposition (PLD)\cite{Ohtomo2002,Hwang2012,Nechache2014}.  One prominent experiment was the chemical synthesis of [KNbO$_{3}$]$_{1-x}$[BaNi$_{1/2}$Nb$_{1/2}$O$_{3-\delta}$]$_{x}$, with $x\approx$ {\it 0.1}, whereby the large band gap of KNbO$_{3}$ (ie. {\it 3.8 } $eV$) is reduced to {\it 1.39} $eV$; though the polarization is reduced from {\it 0.55}$C m^{-2}$ to {\it 0.19} $C m^{-2}$, the  photocurrent is still $\sim$ {\it 50} times that of the classic ferroelectric (Pb,La)(Zr,Ti)O$_{3}$\cite{Grinberg2013}.  Another example with A=A$^\prime$ is Bi$_{2}$FeCrO$_{6}$,  resulting in a band gap of {\it 1.43 - 2.51} $eV$, depending on cation ordering, and a  $\sim${\it 8}$\%$ power conversion efficiency (PCE)\cite{Nechache2014}. However, there is also a significant reduction of ferroelectric polarization, which is detrimental for photovoltaic performance. The preceding examples illustrate the potential of enhancing photovoltaic performance via band gap reduction, though the ferroelectric polarization was excessively suppressed in these cases, which likely degrades the PCE. Therefore, first-principles approaches can be a powerful avenue to more broadly explore this vast phase space, and design low band gap materials which do retain large ferroelectric polarization with strong hybridization.

Engineering double perovskites from first-principles has received some attention within the literature. Previous studies have explored BB$^\prime$ combinations resulting in  d$_{0}$/d$_{10}$ valence states, but these scenarios always result in large band gaps\cite{Takagi2010,Takagi2010a,Roy2011}. Expanding to Mott physics, $d_{n}$/$d_{m}$ scenarios have been successful in attaining low band gaps\cite{Singh2008} but experimental results reveal relatively  small polarizations (ie. {\it 0.2 }$C m^{-2}$ ) and  the origin of ferroelectricity is AA$^\prime$ size mismatch which is not as favorable in terms of hybridization\cite{Li2010}.  Another promising direction is BB$^\prime$ combinations resulting in $d_0$/$d_n$ valencies, which may result in low band gaps\cite{Chen2013}. However, it has been suggested that the nature of the BB$^\prime$ ordering will be critical to realizing a polarization associated with the $d_0$ ion\cite{Knapp2006}, with rock salt ordering being highly unfavorable. This argument is consistent with the lack of polarization in rock salt ordered double perovskite A$_{2}$MWO$_{6}$ ($M$=Ni,Mn,Cu,etc) as determined in experiment\cite{Muoz2002,Azad2002,Iwanaga2000,Khattak1976}.Therefore, it is desirable to pursue layered ordering of BB$^\prime$. Furthermore, octahedral tilting competes with the second order Jahn-Teller distortion associated with the $d_0$ site\cite{Zhong1995}, and having a layered ordering of AA$^\prime$ with large size mismatch can frustrate octahedral tilting\cite{Abakumov2013}.

In this letter, we design a new class of materials where we utilize all of the aforementioned mechanisms simultaneously to achieve a low electronic band gap and a large ferroelectric polarization with strong hybridization. We employ at least one A-type ion with lone pair electrons (ie. Bi$^{3+}$ or Pb$^{2+}$). In order to maximize AA$^\prime$ size mismatch, the A$^\prime$ ion is chosen to have a different valence than the A-type. In particular, Bi$^{3+}$ is paired with Ba$^{2+}$ while Pb$^{2+}$ is paired with La$^{3+}$, where the selection was dictated by maximizing the size of the A-type ion in the respective valence state. We realize full charge transfer with select BB$^\prime$ pairs having valence d$_{n}$/d$_{0}$, which results in a low band gap. Given our AA$^\prime$ selections, nominal valence counting dictates (BB$^\prime$)$^{+7}$, and one of these ions should be $d_0$.  Therefore, we select vanadium as the $d_0$ ion, which must exist in the 5$^{+}$ valence, in conjunction with various B ions in the 2$^{+}$ valence. Vanadium is chosen over, for example, Ti$^{4+}$ or Nb$^{5+}$ for several reasons. First, V$^{5+}$ is smaller than Nb$^{5+}$ and Ti$^{4+}$,\cite{Shannon1976} which will lead to a larger polarization. Second, having a $d_{0}$ B$^\prime$$^{5+}$ion dictates a nominal B$^{2+}$-type ion that is more susceptible to being a Mott insulator for the later transition metals, which are necessary in order to ensure full charge transfer (ie.  sufficient electronegativity differences). Therefore, for B-type ions we choose Ni$^{2+}$ and Cu$^{2+}$. These physically motivated choices are permuted to search for optimal properties, and we additionally explore the validity of these underlying rules.  A final condition is that we restrict this study to layered ordering of AA$^\prime$ and BB$^\prime$ ions, due to the potential for realization via layer-by-layer growth, such as MBE and PLD. We begin with a detailed exploration of BaBiCuVO$_6$, which will serve as our prototype, and then generalize to other possible permutations.

\section{Method}

We utilized density functional theory calculations implemented in the Vienna ab initio simulation program (VASP)\cite{Kresse1993, Kresse1994, Kresse1996a, Kresse1996}, which employs a plane-wave basis and projector-augmented wave (PAW) potentials\cite{Blochl1994, Kresse1999}.  Correlation effects on the 3$d$ orbitals are treated within the local spin density approximation (LSDA) + $U$ method\cite{Anisimov1997}.  The LSDA + $U$ method requires the definition of the on site Coulomb repulsion {\it U}, and we do not include an on site exchange interaction {\it J} given that this is accounted for within LSDA\cite{Park2015}.  We explore a wide of range of values for $U$, given that this is still an open methodological problem.  All results were obtained using $6\times6\times6$ Monkhorst - Pack k-point mesh centered at $\Gamma$ and a {\it 500 }eV plane wave cutoff\cite{Monkhorst1976}.  The tetrahedron method is used for Brillouin zone integration\cite{Blochl1994a}.  In order to find the ground state structure, we have explored several octahedral rotation patterns and computed phonon dispersion curves. Phonons were computed  using the PHONOPY code and a  $2\times2\times1$ supercell with forces obtained from self-consistent total energy calculations using the finite displacement method.\cite{Togo2008} The lattice parameters and atomic positions were relaxed at the LSDA+$U$ levels until the total energy changed by less than {\it 10$^{-5}$ } eV per unit cell and the residual force were smaller than {\it 0.01} eV per $\AA$.  Born effective charges  and the ferroelectric  polarization were computed within the modern theory of polarization\cite{King-smith1993, Resta1994}.  The endpoints of the ferroelectric path are defined using a {\it 180}$^{\circ}$ in-plane rotation centered at the vanadium atom.  Nudged elastic band is performed to determine energy barrier of the ferroelectric path\cite{Henkelman2000}.  Ionic Radii are given by Shannon for different coordination and charge states\cite{Shannon1976}.  All images of crystal structures were generated using VESTA code\cite{Momma2008}.  

\section{Results}

\subsection{Reference structures and spontaneously broken symmetries}

In order to elucidate the origin and nature of the ferroelectric polarization in layered AA$^\prime$BB$^\prime$O$_6$, we consider three reference structures.  The first reference structure, referred to as $R_{o}$, consists of placing the atoms on the ideal perovskite lattice and then allowing full relaxations of the lattice parameters while only relaxing the atoms in the z-direction (space group $Cmm2$, see Figure 1a). The second reference structure, $R_{\vec{q}=0}$, corresponds to fully relaxing $R_o$ with respect to  lattice parameters and all internal coordinates within the unit cell (space group $Cm$, not pictured). The third reference structure,  $R_{\vec{q}=\pi,\pi,0}$, corresponds to creating a supercell with lattice vectors $\bm{(1/\sqrt{2}, 1/\sqrt{2}, 0), (1/\sqrt{2}, -1/\sqrt{2}, 0)}$ and $\bm{ (0, 0, 1) }$ in terms of $R_o$, and then fully relaxing lattice parameters and all internal coordinates (space group $Pc$, see Figure 1b).  These structures allow one to determine the role of $\vec{q}=0$ relaxations and octahedral rotations. It should be noted that even the highest symmetry reference $R_o$ contains a static polarization in the z-direction, though an in-plane ferroelectric polarization can still be formed by spontaneously breaking $C_{4v}$ symmetry via the destruction of a mirror plane upon degenerating to $R_{\vec{q}=0}$ or $R_{\vec{q}=\pi,\pi,0}$.

We begin by examining the R$_o$ structure for BaBiCuVO$_{6}$ (see Figure 1a, 1b).  In this case, we observe a large structural relaxation of both the Ba and Bi towards the Cu-O$_2$ layer, presumably due to the large nominal 5+ charge associated with the V ion. This strong distortion will be seen in all subsequent structures, irrespective of octahedral rotations and other further symmetry breaking. While there is a polarization in the $z$-direction associated with this A/A$^\prime$ ordering and the subsequent z-direction distortion, this is not relevant for our interests as this ordering is static and could not practically be switched.  The main question is therefore how symmetry is spontaneously broken relative to this reference.

In figure 2, we have computed phonon spectra for the three reference structures of BaBiCuVO$_{6}$ (see Methods Section).  In the R$_o$ structure there are many soft-modes, including at the $\Gamma$-point, reflecting the unstable nature of this highest symmetry structure (see Figure 2a).  In the R$_{\vec q=0}$ structure, all of the $\Gamma$-point instabilities have been removed as this structure is a fully relaxed degeneration of the R$_o$ structure (see Figure 2b). This effect is pictorially illustrated in the comparison of the planar projections of Figures 2f and 2g. However, there are still strong in-plane instabilities shown in the phonons, which correctly suggests that there is some lower energy tilting pattern of the octahedron.  In the R$_{\vec q=\pi,\pi,0}$ structure, it is demonstrated that all in-plane instabilities have been removed (see Figure 2c), and the resulting structure is illustrated in the planar projections of Figure 2h and 2i. However, the R$_{\vec q=\pi,\pi,0}$ structure still has a small instability in the z-direction of reciprocal space.  By doubling the unit cell in the z-direction and allowing for full atomic relaxations, we have determined that this instability only causes a very small structural distortion that does not substantially affect the physics (ie. the energy gain is $\approx 1meV$ per f.u.). Therefore, R$_{\vec q=\pi,\pi,0}$ is a sufficient approximation for our ground state structure.

Having established how the symmetry degenerates, we now use the Born effective charges to give an approximate layer-resolved ferroelectric polarization, referenced by R$_{o}$.  The main contribution to the polarization comes from large displacements within the Bi-O and V-O$_2$ layers, which arise from the second order Jahn-Teller effect associated with the lone-pair electrons and $d_0$ configuration, respectively.  In the R$_{\vec q=0}$ structure, the polarization is $\bm{\vec P = 0.435 \vec i +  0.237 \vec j }$ Cm$^{-2}$ , demonstrating that a large polarization is formed even in the absence of any possible octahedral tilts. The layer-resolved analysis further provides the polarization of Bi-O layer and V-O$_2$ layer, resulting in $\bm{\vec P= 0.248\vec i + 0.110\vec j }$ Cm$^{-2}$ and $\bm{ \vec P =0.165 \vec i + 0.110 \vec j }$ Cm$^{-2}$, respectively. In the R$_{\vec q=\pi,\pi,0}$ structure, the magnitude of the polarization remains similar though the direction changes yielding $\bm{\vec P = 0.354 \vec i + 0.354 \vec j }$ Cm$^{-2}$, which can further be split into $\bm{\vec P = 0.189 \vec i + 0.189 \vec j }$ Cm$^{-2}$ for the Bi-O layer and $\bm{\vec P = 0.127 \vec i + 0.127 \vec j }$ Cm$^{-2}$ for the V-O$_2$ layer. Therefore, octahedral tilts do perturb the direction of the polarization.  It should be noted that the polarization in the Bi-O layer is aligned in the same direction as the polarization in the V-O$_2$ layer, giving rise to a very large in-plane polarization.  The possibility of antialigning the Bi-O and V-O$_2$ layers was explored, but this could not be stabilized in our calculations.

\begin{figure} \label{structure}
    \includegraphics[width=80mm]{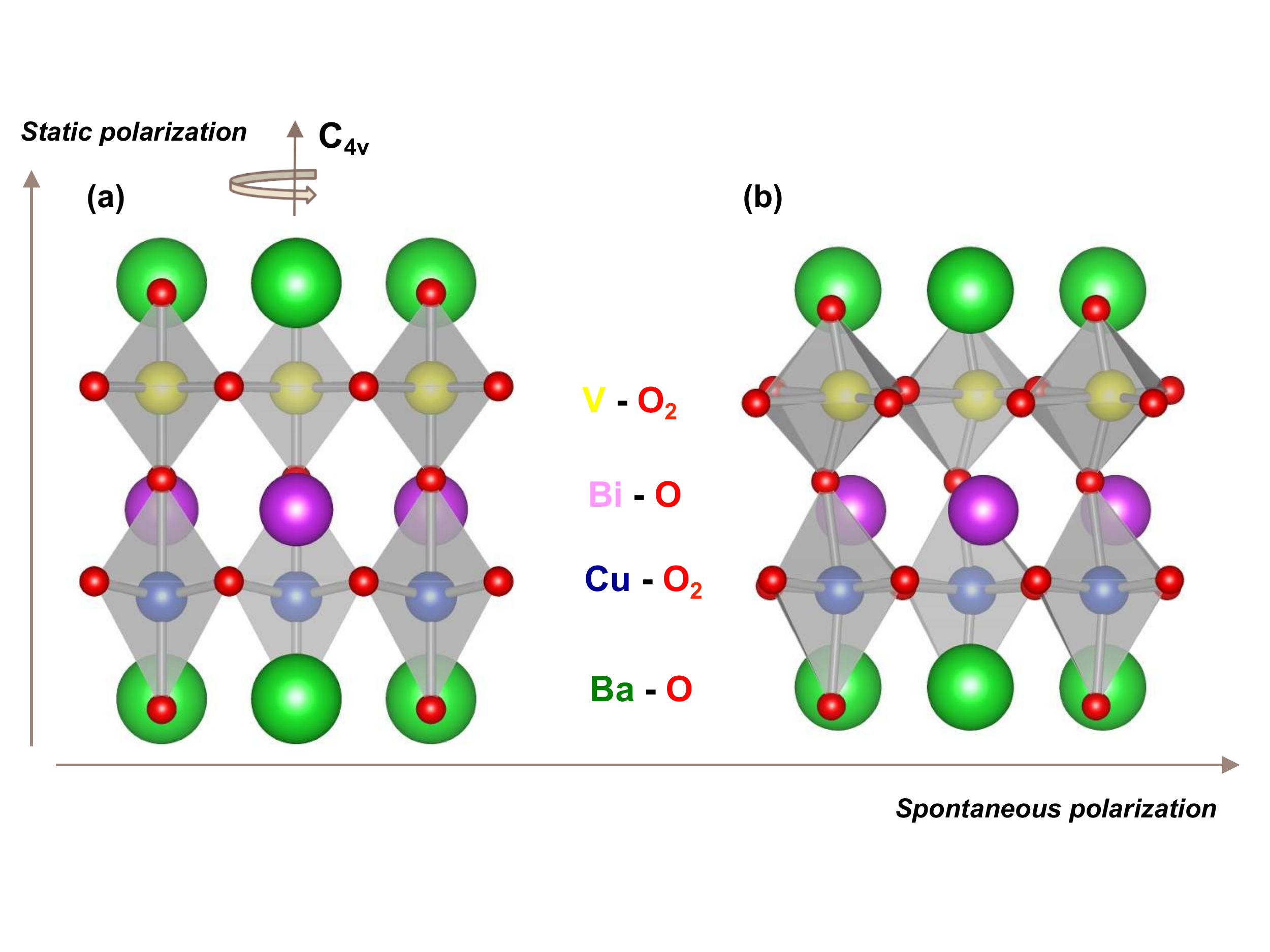}
\caption{{\bf $|$ Reference structures for layered BaBiCuVO$_6$. } {\bf (a)\hspace{1mm} }  Reference structure R$_o$, where the atoms have been placed on an ideal perovskite lattice and subsequently re    laxed only in the z-direction. As illustrated, this structure retains  C$_{4v}$ symmetry, though it is unstable. {\bf (b)\hspace{1mm} } Reference structure R$_{\vec{q}=\pi,\pi,0}$, which allows for ful    l relaxation of all  $\vec{q}=(\pi,\pi,0)$ distortions and lattice strains. }
\end{figure}

\begin{figure}
    \includegraphics[width=80mm]{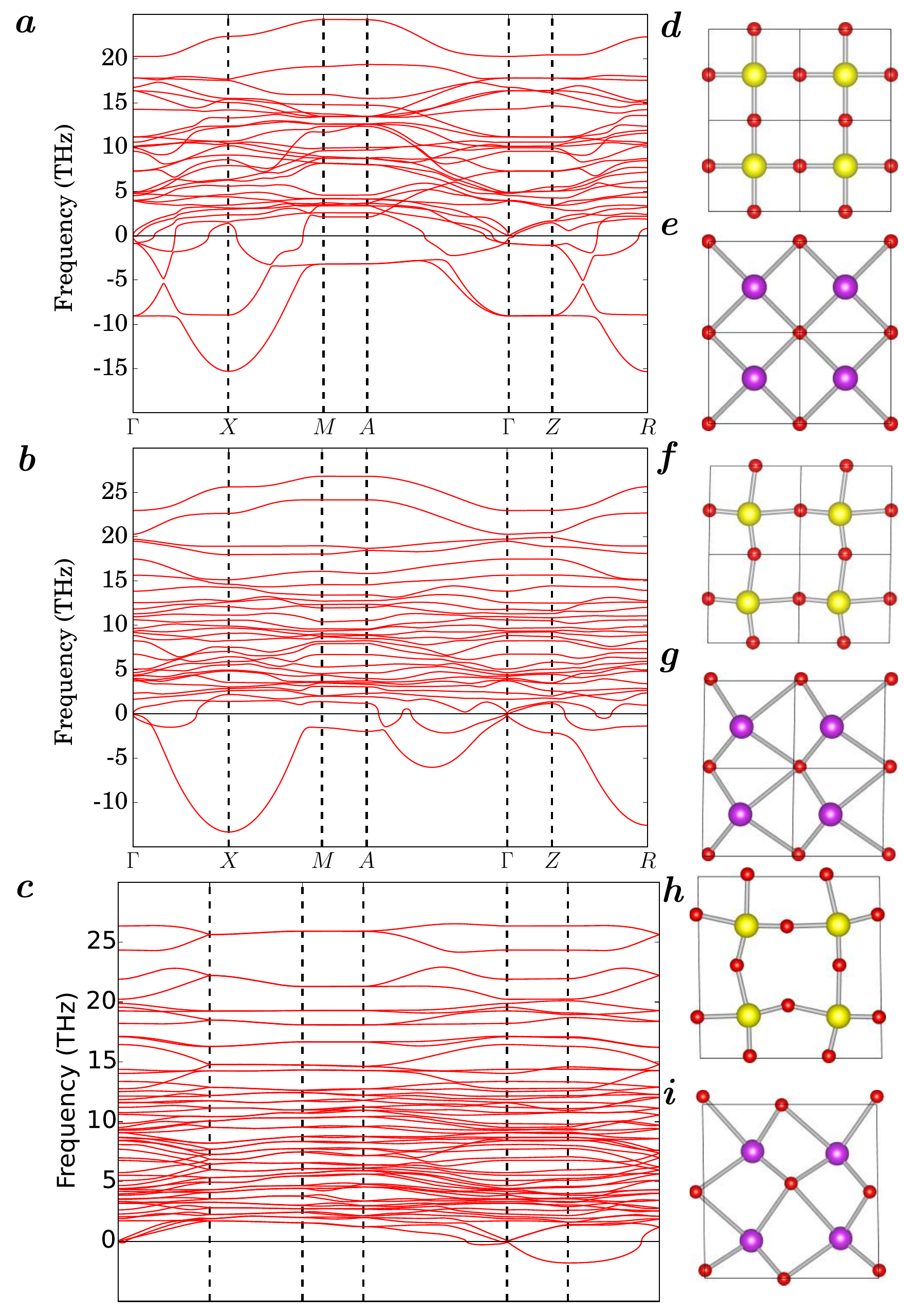} 
\label{phonons}
\caption{
{\bf $|$ Phonons of successive reference structures  }
The phonons of reference structures for BaBiCuVO$_{6}$ are presented in panel {\bf(a)} ,{\bf(b)}, {\bf(c)} corresponding to R$_o$, R$_{\vec{q}=0}$, and R$_{\vec{q}=\pi,\pi,0}$, respectively. Correspond    ing in-plane projections of the Bi-O and V-O layers are given in panels {\bf(d)} ,{\bf(e)} for R$_o$, panels {\bf (f)}, {\bf (g)} for R$_{\vec{q}=0}$, and panels {\bf (h)}, {\bf (i)} for R$_{\vec{q}=\pi,\pi,0}$. In successive order, the polarization is $\bm{0}$, $\bm{0.435 \vec i + 0.237 \vec j }$ $Cm^{-2}$, and $\bm{ 0.354 \vec i +  0.354 \vec j }$ $Cm^{-2}$.  }
\end{figure}

\subsection{Energy barriers and polarization}

\begin{figure}
\includegraphics[width=85mm]{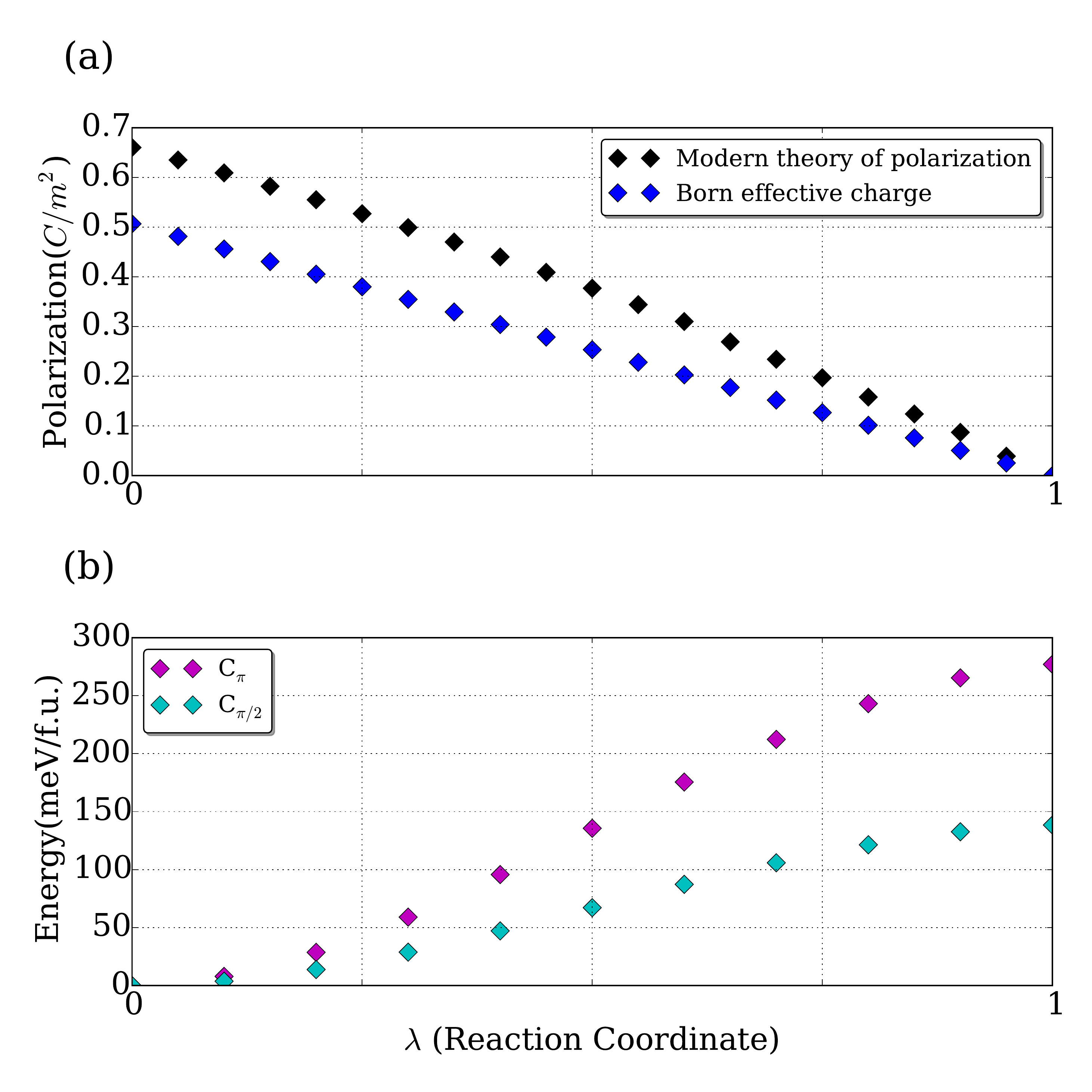}
\caption{{\bf Energy barrier and ferroelectric moment $|$ } {\bf (a)\hspace{1mm} } Total polarization P, {\bf (b)\hspace{1mm} } total energy profile E$_{tot}$ as a function of the polar distortion characterized by $\lambda$ (ie.  $\lambda = 0$ denotes the ferroelectric  phase and $\lambda = 1$ is the midpoint). The energy barrier is obtained via the nudged elastic band method and polarization is calculated through both the modern theory of polarization and Born effective charge analysis of fully distorted structure}
\end{figure}

In order to achieve a precise understanding of the ferroelectricity, we employed the modern theory of polarization to study the ferroelectric moment of BaBiCuVO$_6$ in more detail. The modern theory of polarization has theoretically reproduced the precise magnitude of the ferroelectric moment in many systems\cite{King-smith1993,Resta1994}. In our system, spontaneous breaking of $C_{4v}$ symmetry inherently provides disparate paths between symmetry equivalent minima, and we explore the paths defined by the R$_{\vec q=\pi,\pi,0}$  structure rotated by C$_{\pi}$  and C$_{\pi/2}$, respectively. To begin with, the path defined by C$_{\pi}$ is computed using the  modern theory of polarization yielding 0.65 {\it $C m^{-2}$} (see Figure 3a).  This value is in the vicinity of the polarization of BiFeO$_{3}$\cite{Neaton2005}, which is {\it 0.8 $C m^{-2}$}. Additionally, the corresponding  energy barrier is calculated to be 277 meV per transition metal ion (see Figure 3b). Alternatively, the path defined by C$_{\pi/2}$ results in half of the energy barrier, and this could lead to enhanced switching speeds\cite{Xu2015}. These energy barriers for BaBiCuVO$_{6}$ are reasonable in the sense that BiFeO$_{3}$ has a theoretical energy barrier of {\it 427} $meV per f.u.$ (ie. within Generalized gradient approximation + U) yet the ferroelectric moment is experimentally switchable\cite{Ravindran2006}.

The differences in total polarization between the Berry-phase approach and Born effective charge analysis reflects the presence of anomalous dynamical charges\cite{Ghosez1998} in BaBiCuVO$_6$. In other words, it highlights the covalent character of the atomic bonds and, in turn, of the electronic structure of BaBiCuVO$_6$, which is also seen in  BaTiO$_3$\cite{Ghosez1998}.  Additionally, the born effective charges of active ions in the R$_{o}$ structure exhibit a striking deviation  from nominal charges, which clearly demonstrates a highly covalent character. More specifically, using the Born effective charges from the R$_{o}$ structure yields a polarization of {\it 0.81 $C m^{-2}$}, while using the Born effective charges from the R$_{\vec{q}=\pi,\pi,0}$ structure yields {\it 0.54 $C m^{-2}$}. The average of those two values is similar to the value from the modern theory of polarization, {\it 0.65 $C m^{-2}$}.  These observations are a clear demonstration of the importance of hybridization effects. 

\subsection{Electronic structure and magnitude of band gap}

\begin{figure}
\includegraphics[width=85mm, height=230mm ]{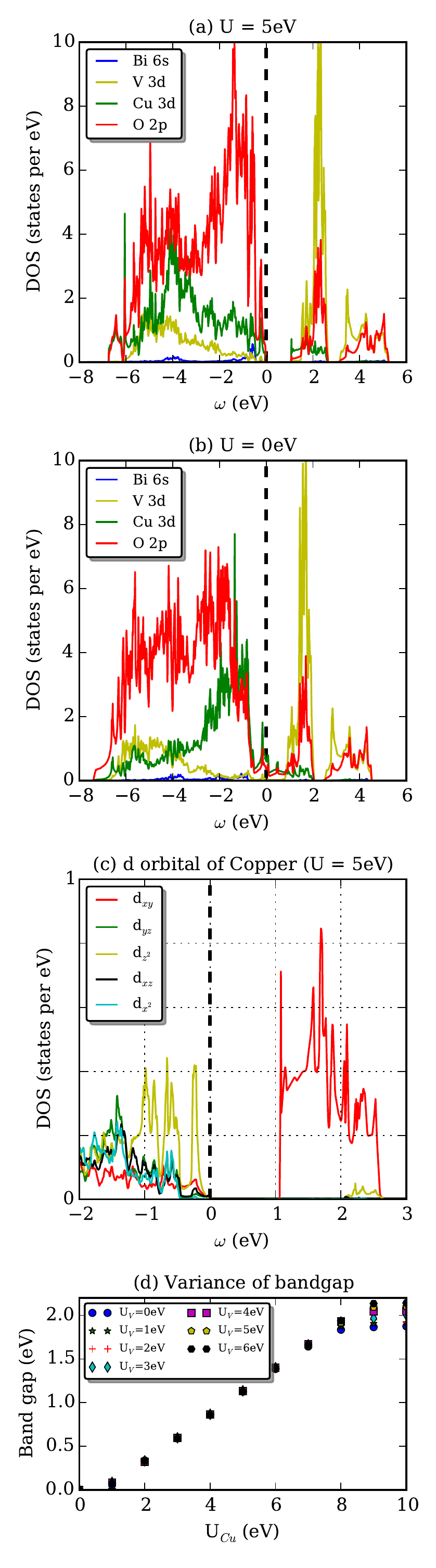}
\caption{{\bf $|$ Electronic structure }  Atomic resolved density of states of BaBiCuVO$_{6}$ in the R$_{\vec q=\pi,\pi,0}$ structure for  {\bf  (a) \hspace{1mm}  } {\it U$_{V}$ = 0 }and {\it U$_{Cu}$ = 5} ; {\bf (b) \hspace{1mm} } {\it U$_V$ = 0 } and {\it U$_{Cu}$ = 0 eV } ; {\bf (c)\hspace{1mm} } projection onto copper d-orbitals; Fermi energy is set to be {\it 0};  {\bf (d)\hspace{1mm} } The variance of band gap depending on {\it U$_{V}$} and {\it U$_{Cu}$}.  } 
\end{figure}

Here we explore the basic electronic structure of  BaBiCuVO$_{6}$, which follows the simple rules outlined in the introduction.  More specifically, V nominally donates its sole $d$-electron to Cu, analogous to what was observed for Ti in La$_2$NiTiO$_6$\cite{Chen2013}.  We begin by illustrating this via density functional theory (DFT)+U calculations, where we use a Hubbard $U$ of {\it 5 eV} on the copper site, consistent with previous studies \cite{Onsten2007}, and we conservatively use $U=${\it 0} on the V site. The effect of the on-site Hubbard $U$ will be explored in detail below. The standard fully localized limit double counting\cite{Anisimov1993,Czyzyk1994} is used, and no on-site exchange is employed given that we are using a spin-dependent density functional\cite{Park2015}. The resulting density of states (DOS) for the ground state antiferromagnetic spin ordering yields a band gap of {\it 1.1} $eV$ (see Figure 4a).  The atom resolved DOS for V exhibits a strong peak above the gap indicating a complete nominal charge transfer, which is corroborated by zero magnetic moment on the V atom. However, there is a substantial orbital occupation of $\approx$ {\it 4} electrons  for the V $d$-orbitals due to strong hybridization with the oxygen, as illustrated by the substantial DOS projection of V towards the bottom of the oxygen bands. Below we demonstrate that this results in the band gap being very insensitive to the value of $U$ on the V site. Cu has a magnetic moment of {\it  0.5} $\mu_B$, which is substantially reduced from the nominal value of {\it 1} $\mu_B$.  Projecting the DOS onto the Cu-3d states reveals states above the gap, which are driven by the on-site Coulomb repulsion $U$ (see Figure 4a, 4b). This scenario is analogous to the cuprates, where there is a substantial crystal field that breaks $E_{g}$ symmetry and results in an effective single band Hubbard model which yields an antiferromagnetic Mott insulator. This is clearly illustrated by the orbitally resolved DOS presented in Figure 4c, where $d_{xy}$ character is seen both above and below the gap while $d_{z^2}$ is nearly completely filled.

The precise value of the on-site Coulomb repulsion $U$ is still an open problem, and the DFT+U method is crude in the sense that it is a Hartree approximation to the dynamical mean field theory(DMFT) impurity problem within the DFT+DMFT formalism.  Both of these issues warrant a careful exploration of how the results are affected by changes in $U$.  Therefore, we compute the band gap of the system as a function of $U$ for vanadium and copper, exploring essentially all feasible values (ie. {\it 0 - 6}$eV$, {\it 0 - 10} $eV$ for vanadium and copper, respectively; See Figure 4d). The $U$ of vanadium has very little effect due to the large hybridization with the oxygen. Neglecting exchange, the orbital dependent  term within DFT+U that is added to the Kohn-Sham potential has the form $U(1/2-n_i)$\cite{Anisimov1997}, such that orbitals near half filling will not be substantially perturbed unless $U$ is sufficiently large relative to the hybridization.  Alternatively, the $U$ on Cu is essential for driving the Mott physics and is necessary to open a gap (see Figure 4a, 4b).  Increasing the value of $U$ increases the gap, as expected, though the magnitude of the increase is relatively small due to the strong hybridization of Cu and oxygen, and eventually the band gap largely saturates due to the upper band edge being dominated by V. Even for a relatively large value of $U =$ {\it 6} $eV$, the band gap remains relatively small at {\it 1.35} $eV$.  This scenario should be contrasted to La$_2$TiNiO$_6$, where the gap depended strongly on the $U$ value for Ti\cite{Chen2013} due to a relatively smaller electronegativity difference for Ti-Ni and less Ti-O  hybridization.

\subsection{Generalization to other compounds}
\begingroup
\squeezetable
\begin{table*}[ht]
\centering
\caption{{\bf The summary of planar Ferroelectric Mott $|$ } The electronic gap, energy barrier, (layer-resolved) polarization, and AA$^\prime$ size difference based on Shannon radii\cite{Shannon1976} for AA$^\prime$BB$^\prime$O$_6$ compounds.  The energy barrier is normalized per transition metal. Total polarization is computed by modern theory of polarization.  The asterisk indicates that $U$ was increased to {\it 6.5} $eV$ and {\it 8.0} $eV$ on Ni and Cu, respectively, in order to ensure a gap was maintained over the entire distortion path, which is a necessary condition for the modern theory of polarization. This change has a negligible effect on the other observables.  }
\medskip
{
  \begin{tabular}{|c||c|c|c|c|}
\hline
AA$^\prime$BB$^\prime$O$_6$ & BaBiCuVO$_{6}$ & PbLaCuVO$_{6}$  & BaBiNiVO$_{6}$ & BaLaCuVO$_{6}$ \\
\hline
 Gap (eV)                          &  1.08    &  1.00    &  1.05    &  0.86    \\  
Energy barrier (meV)               &  287     &  226    &   269    &  214     \\  
 Polarization ($C m^{-2}$)          &  0.65    &  0.52$^{*}$   &  0.68$^{*}$    &  0.41$^{*}$    \\  
 A/A$^\prime$size mismatch ($\AA$)  &  0.25    &  0.13    &  0.25    &  0.26    \\  
 Unit Cell Volume ($\AA^{3}$)            &  240.47  &  229.72  & 237.41  &    239.38  \\  
 A-O             ($C m^{-2}$)      &  0.04       &  0.12    &  0.03    &  0.03    \\  
 A$^\prime$-O    ($C m^{-2}$)      &  0.28    &  0.10    &  0.28   &  0.08    \\  
 B-O             ($C m^{-2}$)      &  0.00    &  0.00     &  0.03    &  0.00       \\  
 B$^\prime$-O    ($C m^{-2}$)      &  0.19    &  0.23    &  0.22   &  0.16    \\  

 \hline
  \end{tabular}
}
\end{table*}
\endgroup

While we have illustrated our design principles in the context of BaBiCuVO$_6$, these rules can obviously be used to generate numerous other compounds with similar properties.  To illustrate this point, we have successfully realized three additional candidates in addition to  BaBiCuVO$_6$: PbLaCuVO$_{6}$, BaBiNiVO$_{6}$, and BaLaCuVO$_{6}$ (see Table 1).   We first consider the case  of PbLaCuVO$_{6}$, where we have decreased the A/A$^\prime$ mismatch while retaining one lone-pair cation and kept everything else identical.  As expected\cite{Takagi2010}, the smaller A/A$^\prime$ mismatch results in  a smaller polarization within the lone-pair cation layer, though a nontrivial polarization does form in the La-O layer. Nonetheless, the net effect is an overall decrease in polarization and a large decrease in the energy barrier, which could be beneficial for ease of electrical switching.  Due to the relatively small changes in the Cu-O$_2$ planes, the band gap changes very little.  Another interesting possibility is to retain the Ba/Bi cations and instead exchange Ni in place of Cu. In this case, Ni will result in a 2+ state, which is susceptible to Mott physics, similar to NiO and the scenario outlined in La$_2$TiNiO$_6$\cite{Chen2013}.  We find a similar barrier and polarization to BaBiCuVO$_6$, which is consistent with the only change being associated with the ion exhibiting Mott physics. Additionally, the band gaps are relatively similar, assuming a reasonable value of $U=${\it 5}$eV$ for Ni\cite{Park2012}.  The final scenario we explore corresponds to the removal of lone-pair cations, which is the case of BaLaCuVO$_{6}$. Here the only nontrivial polarization in the A/A$^\prime$ layers is associated with La-O, which is analogous to the  large size mismatch as in the  KNbO$_3$-LiNbO$_3$ system\cite{Bilc2006}. The band gap is slightly reduced relative to BaBiCuVO$_6$ due to diminished tilting in the Cu-O$_2$ layer.

\section{Discussion}
Our proposed  class of planar ferroelectric Mott insulators optimally matches the criteria  for use as a  photo-ferroelectric\cite{Butler2014, Shockley1961}: band gap of $\approx 1eV$, strong ferroelectric polarization and covalent character, and a large phase space of chemical elements such that properties can be tuned.  The low band gap is essential to absorb a significant portion of the solar spectrum. The sizable polarization and strong covalent character are necessary to obtain a substantial shift current\cite{Young2012, Young2012a}, which is believed to be indicative of the photocurrent and hence the power conversion efficiency (PCE).  Our current work has identified four compounds which satisfy our design rules, and there are many more compounds which could be included to form a substantial phase space of possibilities.  Additionally, the fact that our systems are antiferromagnet Mott insulators  could further enhance the potential of these materials in two ways\cite{Manousakis2010}. First, multiple charge carriers through impact ionization suggest the possibility of exceeding the Shockley-Queisser efficiency limit\cite{Shockley1961}. Second, high drift voltage and large diffusion length could be helpful despite a low mobility due to strong correlations.  Antiferromagnetic Mott insulators have demonstrated the possibility of enhancing  ultrafast separation of photo doped carriers before recombination\cite{Eckstein2014}.  Only careful experiments will be able to definitively determine the applicability of Mott insulators in this context due to the complexity of strongly correlated electron systems.

Another potential area of application exists in the context of non-volatile memory.  The photovoltaic output of BiFeO$_3$ was employed as a read-out signal, switching between {\it 0.11 $V$ and -0.23 $V$ } for up to {\it 108} cycles\cite{Guo2013}.  In terms of non-volatile memory devices, an in-plane ferroelectric polarization has several benefits\cite{Yao2005}.  First, in-plane ferroelectrics should be much less sensitive to explicitly broken symmetry induced by a substrate, leading to a more symmetric voltage and current bi-state.  Second, in-plane ferroelectrics may be straightforwardly exploited via the growth of large area thin-films generating a large voltage, which would be difficult to achieve in a thick film out-of-plane ferroelectric. An open circuit voltage of 8 volts is achieved with a small incident intensity of {\it 1} $mW cm^{-2}$ for a thin film device of Pb$_{0.97}$La$_{0.03}$(Zr$_{0.52}$Ti$_{0.48}$O$_{3}$. Thus our proposed class of materials could be multifunctional in the sense that the ferroelectric order is coupled to optical and electronic transport channels. This could lead to optically addressed ferroelectric memory with non-destructive read out\cite{Wu2011}.  It is also worth noting that a large planar ferroelectric moment could assist the formation of charged domain walls, which have their own potential device applications\cite{Oh2015}.

It is worthwhile to compare our systems with the recent resurgence of organometallic halides in the context of photovoltaic, which has also resolved the aforementioned deficiencies of ferroelectric oxide photovoltaic\cite{Butler2014}. In the organometallic halides, variation in hysteresis of the  current-voltage profile under different  scanning conditions implies ferroelectric behavior and exceptionally long carrier diffusion lengths, leading to the hypothesis of  ``ferroelectric highways''.  Recent ab-initio calculations confirm a large polarization and strong covariance in photovoltaicly promising hybrid lead halide perovskites\cite{Frost2014, Zheng2015}, which is consistent with the general view of shift current theory\cite{Young2012}.  Nonetheless, their instability against water due to weak halide bonding and environmental issues associated with lead are still very serious problems\cite{Green2014}, and resolving these deficiencies could be formidable. Therefore, the approach outlined in this paper could be a very valuable path to explore in parallel with the recent trends in organometallics.  

\begin{acknowledgements}
We would like to thank D. Vanderbilt and K. Rabe for valuable discussions. The authors acknowledge funding from the FAME, one of six centers of STARnet, a Semiconductor Research Corporation program sponsored by MARCO and DARPA. The research used resources of the National Energy Research Scientific Computing Center, a DOE Office of Science User Facility supported by the Office of Science of the U. S. Department of Energy under Contract No. DE-AC02-05CH11231
\end{acknowledgements}

\bibliography{reference}

\end{document}

\newcommand\xmas{\mbox{$x_{\rm mas}$}}
\newcommand\rmas{\mbox{$r_{\rm mas}$}}
\newcommand\Xmas{\mbox{$X_{\rm mas}$}}
\newcommand\Hinf{\mbox{${\rm H}_{\rm inf}$}}
\newcommand\Omegainf{\mbox{$\Omega_{\rm inf}$}}
\newcommand\Omegabg{\mbox{$\Omega_{\rm bg}$}}
\newcommand\Hbg{\mbox{${\rm H}_{\rm bg}$}}
\newcommand\rhobg{\mbox{$\rho_{\rm bg}$}}
\newcommand\rhorbg{\mbox{$\rho^{\rm bg}_{\rm r}$}}
\newcommand\rhoinf{\mbox{$\rho_{\rm inf}$}}
\newcommand\kinf{\mbox{$k_{\rm inf}$}}
\newcommand\kbg{\mbox{$k_{\rm bg}$}}
\newcommand\ainf{\mbox{$a_{\rm inf}$}}
\newcommand\abg{\mbox{$a_{\rm bg}$}}
\newcommand\xpbg{\mbox{$x_{\rm p}^{\rm bg}$}}
\newcommand\xpinf{\mbox{$x_{\rm p}^{\rm inf}$}}
\newcommand\xmasbg{\mbox{$x_{\rm mas}^{\rm bg}$}}
\newcommand\xmasinf{\mbox{$x_{\rm mas}^{\rm inf}$}}
\newcommand\thetainf{\mbox{$\theta^{\rm inf}$}}
\newcommand\thetabg{\mbox{$\theta^{\rm bg}$}}
\newcommand\xHbg{\mbox{$1/{\rm H}_{\rm bg}$}}
\newcommand\xHinf{\mbox{$1/{\rm H}_{\rm inf}$}}

\title{Supplementary Information  for 

'New class of planar ferroelectric Mott insulators via first principles design'}

\author{Chanul Kim}
\author{Hyowon Park } 
\author{Chris A. Marianetti}

\maketitle

{\bf 1. GROUND STATE STRUCTURES} 

\begin{figure}[h]
\includegraphics[width=\linewidth,height=\textheight,keepaspectratio]{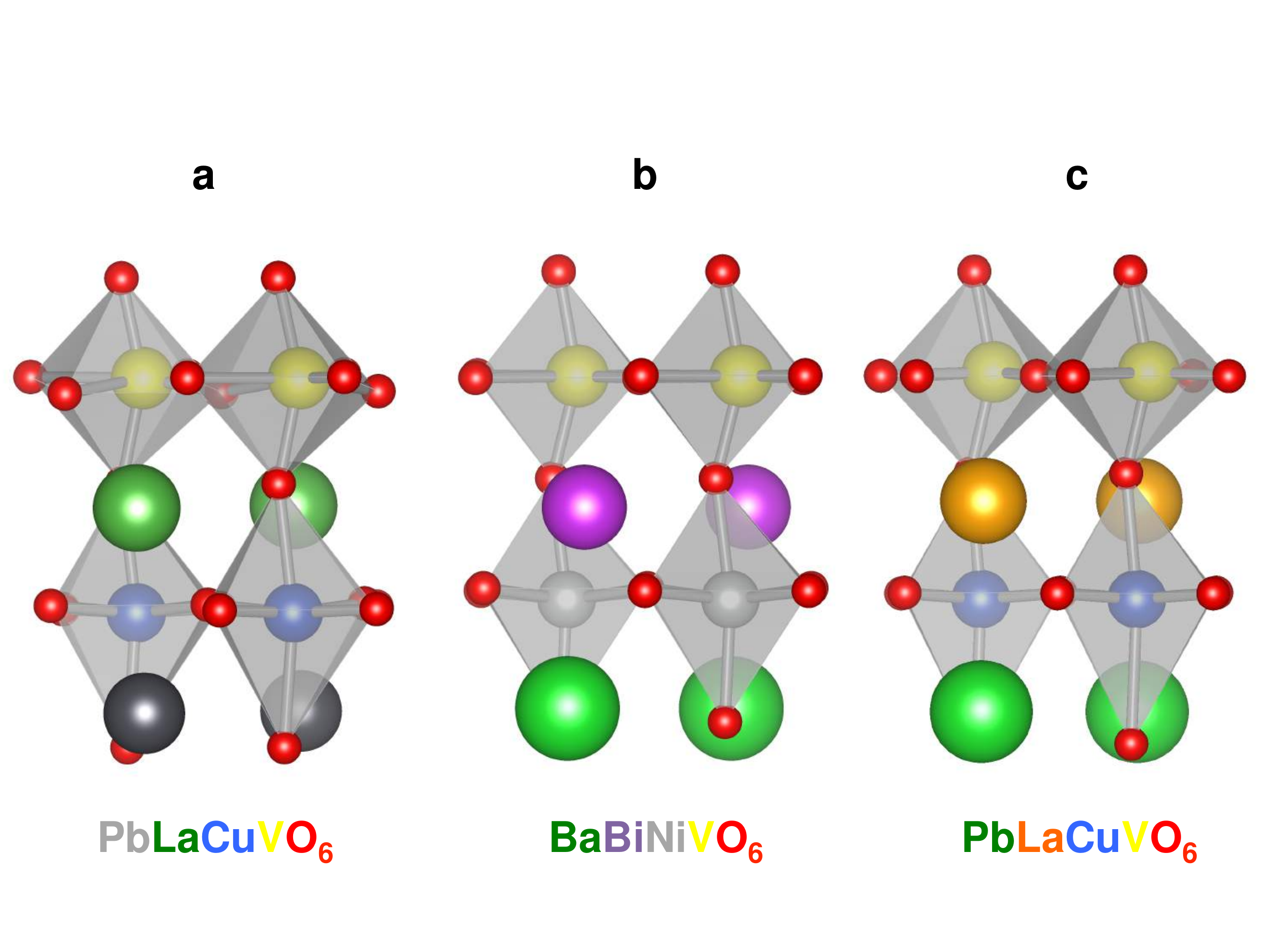}
\caption{{\bf $|$ Fully relaxed structure of AA$^\prime$BB$^\prime$O$_6$} {\bf (a)\hspace{1mm} } PbLaCuVO$_6$ {\bf (b)\hspace{1mm} } BaBiNiV$_6$ {\bf (c)\hspace{1mm} } BaLaCuVO$_6$
; The colors of chemicla formula and structure correspond the species of atoms}
\end{figure}

\newpage

{\bf 2. ATOMIC COORDINATES, LATTICE PARAMETERS and BORN EFFECTIVE CHARGES}

\begin{table}[h]
\caption{Born Effective Charges  and reduced coordinates of BaBiCuVO$_6$ in the orthorhombic 20-atoms cell.}
    \begin{center}
           \begin{tabular}{*{20}{c}}
           \hline \hline 
            \\ [-7 pt]       
            & & &   & &   & & &  a = 5.387 & &  b = 5.457 & & c= 8.180 \\      
Atom  &  Z$^*_{xx}$ &  &   Z$^*_{yy}$ &  &   Z$^*_{zz}$  &  &  &    $x$   &  &     $y$      &  &     $z$      \\
 Ba   &  2.834   &  &   2.809    &  &   3.564       &  &  & 0.25514276 &  &     0.24534037    &  &   0.26140472   \\
 Ba   &  2.834   &  &   2.809    &  &   3.564       &  &  & 0.74485741 &  &     0.74534082    &  &   0.26140470   \\
 Bi   &  5.054   &  &   4.961    &  &   3.888       &  &  & 0.26351321 &  &     0.19642385    &  &   0.70713054   \\
 Bi   &  5.054   &  &   4.962    &  &   3.888       &  &  & 0.73648696 &  &     0.69642405    &  &   0.70713040   \\
  V   &  6.518   &  &   4.661    &  &   5.338       &  &  & 0.24516770 &  &     0.71660716    &  &  -0.00331922   \\
 V    &  6.509   &  &   4.664    &  &   5.344       &  &  & 0.75483237 &  &     0.21660449    &  &  -0.00331904   \\
  C   &  2.089   &  &   2.151    &  &   1.791       &  &  & 0.24564854 &  &     0.73988309    &  &   0.50548329   \\
 Cu   &  2.089   &  &   2.151    &  &   1.791       &  &  & 0.75435144 &  &     0.23988311    &  &   0.50548335   \\
 O    &  -3.983  &  &    -2.191  &  &     -1.333    &  &  & 0.02899101 &  &     0.05405394    &  &  -0.02243997   \\
 O    &  -3.983  &  &    -2.191  &  &     -1.333    &  &  & -0.02899188&  &      0.55405366   &  &   -0.02243998  \\
 O    &  -3.203  &  &    -2.698  &  &     -1.18     &  &  & 0.53612190 &  &    -0.01001588    &  &   0.00951961   \\
 O    &  -3.203  &  &    -2.698  &  &     -1.18     &  &  & 0.46387841 &  &     0.48998308    &  &   0.00951949   \\
  O   &  -1.830  &  &   -1.913   &  &    -3.98      &  &  & 0.22966860 &  &     0.77063504    &  &   0.21373885   \\
 O    &  -1.830  &  &   -1.912   &  &    -3.98      &  &  & 0.77033186 &  &     0.27063572    &  &   0.21373930   \\
 O    &  -2.573  &  &    -2.584  &  &     -2.305    &  &  & 0.00094419 &  &    -0.00745141    &  &   0.53173795   \\
 O    &  -2.572  &  &    -2.586  &  &     -2.302    &  &  & -0.00094471&  &      0.49254838   &  &    0.53173805  \\
 O    &  -2.562  &  &    -2.546  &  &     -2.502    &  &  & 0.49446762 &  &     0.00155895    &  &   0.51652627   \\
 O    &  -2.563  &  &    -2.547  &  &     -2.503    &  &  & 0.50553274 &  &     0.50155854    &  &   0.51652629   \\
 O    &  -2.347  &  &    -2.655  &  &     -3.275    &  &  & 0.29830525 &  &     0.79296632    &  &   0.78021798   \\
 O    &  -2.347  &  &    -2.654  &  &     -3.279    &  &  & 0.70169458 &  &     0.29296668    &  &   0.78021739   \\
           \hline \hline 
           \\ [-7 pt]
      \end{tabular} 
     \end{center}
\label{tab:Born}
\end{table}

\begin{table}[h]
    \caption{Born Effective Charges  and reduced coordinates of PbLaCuVO$_{6}$ in the orthorhombic 20-atoms cell.}
    \begin{center}
           \begin{tabular}{*{20}{c}}
           \hline \hline 
            \\ [-7 pt]       
            & & &   & &   & & &  a = 5.387 & &  b = 5.309  &  & c= 8.053 \\      
Atom  &  Z$^*_{xx}$ &  &   Z$^*_{yy}$ &  &   Z$^*_{zz}$  &  &  &    $x$   &  &     $y$      &  &     $z$      \\
Pb   &  3.575  &  &    3.591  &  &   4.189     &  &  &   0.22158575  &  &   0.24809749    &  &    0.28782581  \\
Pb   &  3.573  &  &    3.592  &  &   4.2       &  &  &   0.72158674  &  &   0.75190224    &  &    0.28782594  \\
La   &  4.343  &  &    4.246  &  &   5.033     &  &  &   0.23719897  &  &   0.25386950    &  &    0.72602372  \\
La   &  4.34  &  &    4.247  &  &   5.031      &  &  &   0.73721805  &  &   0.74613092    &  &    0.72602414  \\
V    &  4.804  &  &    6.169  &  &   5.659     &  &  &   0.21381777  &  &   0.74310300    &  &   -0.00020160  \\
V    &  4.803  &  &    6.167  &  &   5.655     &  &  &   0.71386053  &  &   0.25689634    &  &   -0.00023513  \\
Cu   &  2.122  &  &    2.035  &  &   1.571     &  &  &   0.24320996  &  &   0.74998738    &  &    0.49966060  \\
Cu   &  2.121  &  &    2.034  &  &   1.571     &  &  &   0.74321866  &  &   0.25001467    &  &    0.49966488  \\
 O   &  -3.074  &  &    -3.14  &  &   -1.511   &  &  &   -0.03795712 &  &   -0.05741095   &  &    -0.03023650 \\
O    &  -2.14  &  &    -3.751  &  &   -1.419   &  &  &   0.07014963  &  &   0.45086863    &  &    0.00336712  \\
O    &  -3.075  &  &    -3.14  &  &   -1.51    &  &  &   0.46205099  &  &   0.05742094    &  &   -0.03023610  \\
O    &  -2.14  &  &    -3.753  &  &   -1.419   &  &  &   0.57015360  &  &   0.54911884    &  &    0.00336638  \\
O    &  -2.235  &  &    -1.962  &  &   -3.785  &  &  &   0.27643088  &  &   0.78697212    &  &    0.21341166  \\
O    &  -2.234  &  &    -1.961  &  &   -3.788  &  &  &   0.77641671  &  &   0.21303226    &  &    0.21340697  \\
O    &  -2.481  &  &    -2.657  &  &   -2.98   &  &  &   0.01499215  &  &   0.01902553    &  &    0.51728328  \\
 O   &  -2.598  &  &    -2.309  &  &   -3.056  &  &  &   -0.02317921 &  &    0.51912615   &  &     0.50752395 \\
O    &  -2.48  &  &    -2.657  &  &   -2.979   &  &  &   0.51498266  &  &  -0.01903757    &  &    0.51728362  \\
O    &  -2.598  &  &    -2.309  &  &   -3.056  &  &  &   0.47680972  &  &   0.48088314    &  &    0.50752341  \\
O    &  -2.305  &  &    -2.215  &  &   -3.696  &  &  &   0.28373604  &  &   0.72132392    &  &    0.77533943  \\
O    &  -2.304  &  &    -2.215  &  &   -3.700  &  &  &   0.78371752  &  &   0.27867545    &  &    0.77537846  \\
            \\ [-7 pt]
           \hline \hline 
      \end{tabular} 
     \end{center}
\label{tab:Born}
\end{table}

\newpage 

\begin{table}[h]
    \caption{Born Effective Charges  and reduced coordinates of BaBiNiVO$_{6}$  in the orthorhombic 20-atoms cell.}
    \begin{center}
           \begin{tabular}{*{20}{c}}
           \hline \hline 
            \\ [-7 pt]       
            & & &   & &   & & &  a = 5.583 & &  b = 5.550 & & c= 7.727 \\      
Atom  &  Z$^*_{xx}$ &  &   Z$^*_{yy}$ &  &   Z$^*_{zz}$  &  &  &    $x$   &  &     $y$      &  &     $z$      \\
  Ba   &  2.839  &  &    2.912  &  &   3.441      &  &  &   0.24717643 &  &    0.24993781  &  &   0.25911660  \\
  Ba   &  2.839  &  &    2.912  &  &   3.442      &  &  &   0.74717600 &  &    0.75006205  &  &   0.25911615  \\
  Bi   &  5.107  &  &    4.93  &  &   3.626       &  &  &   0.19372302 &  &    0.25087924  &  &   0.70519973  \\
  Bi   &  5.107  &  &    4.93  &  &   3.627       &  &  &   0.69372264 &  &    0.74912477  &  &   0.70519960  \\
  V    &  4.876  &  &    6.382  &  &   5.903      &  &  &   0.21414132 &  &    0.75021351  &  &  -0.00276106  \\
  V    &  4.875  &  &    6.381  &  &   5.903      &  &  &   0.71413691 &  &    0.24978739  &  &  -0.00276153  \\
  Ni   &  1.836  &  &    1.798  &  &   2.312      &  &  &   0.24720605 &  &    0.75028751  &  &   0.50068445  \\
  Ni   &  1.835  &  &    1.798  &  &   2.312      &  &  &   0.74720523 &  &    0.24971204  &  &   0.50068463  \\
  O    &  -2.551  &  &    -3.609  &  &   -1.265   &  &  &   0.01634107 &  &   -0.00674227  &  &  -0.00141176  \\
  O    &  -2.502  &  &    -3.663  &  &   -1.289   &  &  &   0.02058760 &  &    0.50323374  &  &  -0.00626253  \\
  O    &  -2.551  &  &    -3.609  &  &   -1.265   &  &  &   0.51633993 &  &    0.00674113  &  &  -0.00140971  \\
  O    &  -2.502  &  &    -3.663  &  &   -1.289   &  &  &   0.52058668 &  &    0.49676624  &  &  -0.00626069  \\
  O    &  -1.824  &  &    -1.729  &  &   -4.191   &  &  &   0.26490638 &  &    0.74696919  &  &   0.22990260  \\
  O    &  -1.822  &  &    -1.729  &  &   -4.191   &  &  &   0.76491031 &  &    0.25303069  &  &   0.22990495  \\
  O    &  -2.568  &  &    -2.593  &  &   -2.414   &  &  &   0.00826344 &  &    0.00668110  &  &   0.52036953  \\
   O   &  -2.666  &  &    -2.608  &  &   -2.356   &  &  &   -0.00031312&  &     0.50182921 &  &    0.52601810 \\
  O    &  -2.569  &  &    -2.593  &  &   -2.413   &  &  &   0.50826473 &  &   -0.00667956  &  &   0.52036855  \\
  O    &  -2.665  &  &    -2.608  &  &   -2.356   &  &  &   0.49968819 &  &    0.49816922  &  &   0.52601719  \\
  O    &  -2.549  &  &    -1.822  &  &   -3.762   &  &  &   0.28796856 &  &    0.75605814  &  &   0.76914401  \\
  O    &  -2.551  &  &    -1.821  &  &   -3.768   &  &  &   0.78796864 &  &    0.24393882  &  &   0.76914124  \\
            \\ [-7 pt]
           \hline \hline 
      \end{tabular} 
     \end{center}
\label{tab:Born}
\end{table}

\begin{table}[h]
    \caption{Born Effective Charges and reduced coordinates of BaLaCuVO$_{6}$ in the orthorhombic 20-atoms cell.}
    \begin{center}
           \begin{tabular}{*{20}{c}}
           \hline \hline 
            \\ [-7 pt]       
            & & &   & &   & & &  a = 5.426 & &  b = 5.364 & & c= 8.225 \\      
Atom  &  Z$^*_{xx}$ &  &   Z$^*_{yy}$ &  &   Z$^*_{zz}$  &  &  &    $x$   &  &     $y$      &  &     $z$      \\
 Ba    &  2.704  &  &    2.712  &  &   3.625      &  &  & 0.25296977 &  & 0.24241584 &  &-0.02357038    \\
 Ba    &  2.705  &  &    2.712  &  &   3.628      &  &  & 0.75294014 &  & 0.74421133 &  &-0.02354220    \\
 La    &  4.34  &  &    4.272  &  &   4.067       &  &  & 0.26138195 &  & 0.24397095 &  & 0.53009295    \\
 La    &  4.341  &  &    4.27  &  &   4.066       &  &  & 0.76141447 &  & 0.74299694 &  & 0.53008170    \\
 V     &  4.486  &  &    5.887  &  &   5.126      &  &  & 0.29296837 &  & 0.75169868 &  & 0.25690697    \\
 V     &  4.481  &  &    5.887  &  &   5.121      &  &  & 0.79300514 &  & 0.23483067 &  & 0.25685807    \\
 Cu    &  2.149  &  &    2.06  &  &   1.491       &  &  & 0.25227739 &  & 0.74328231 &  & 0.73810075    \\
 Cu    &  2.148  &  &    2.06  &  &   1.496       &  &  & 0.75223399 &  & 0.24358012 &  & 0.73811651    \\
 O     &  -1.764  &  &    -1.725  &  &   -3.872   &  &  & 0.23399499 &  & 0.73932052 &  & 0.04451123    \\
 O     &  -1.761  &  &    -1.722  &  &   -3.871   &  &  & 0.73376116 &  & 0.24739132 &  & 0.04448562    \\
  O    &  -2.057  &  &    -3.496  &  &   -1.238   &  &  & -0.06691756&  & -0.05553194&  &  0.26510464   \\
 O     &  -2.872  &  &    -2.937  &  &   -1.227   &  &  & 0.04909435 &  & 0.42648228 &  & 0.26581048    \\
 O     &  -2.06  &  &    -3.496  &  &   -1.237    &  &  & 0.43313158 &  & 0.04201058 &  & 0.26529792    \\
 O     &  -2.87  &  &    -2.938  &  &   -1.227    &  &  & 0.54901425 &  & 0.55995916 &  & 0.26606564    \\
 O     &  -2.125  &  &    -2.06  &  &   -3.117    &  &  & 0.21916248 &  & 0.73796316 &  & 0.47206989    \\
 O     &  -2.121  &  &    -2.06  &  &   -3.12     &  &  & 0.71969351 &  & 0.24841336 &  & 0.47209169    \\
 O     &  -2.418  &  &    -2.276  &  &   -2.41    &  &  & 0.00758795 &  & 0.00039971 &  & 0.72461584    \\
  O    &  -2.414  &  &    -2.418  &  &   -2.429   &  &  & -0.00266743&  &  0.49627268&  &  0.72621931   \\
 O     &  -2.419  &  &    -2.275  &  &   -2.408   &  &  & 0.50755743 &  &-0.01351650 &  & 0.72457668    \\
 O     &  -2.416  &  &    -2.418  &  &   -2.425   &  &  & 0.49739609 &  & 0.49064891 &  & 0.72610664    \\
            \\ [-7 pt]
             \hline \hline
      \end{tabular} 
     \end{center}
\label{tab:Born}
\end{table}

\newpage

{\bf 3. DENSITY OF STATES}

\begin{figure}[h]
\includegraphics[width=\linewidth,height=\textheight,keepaspectratio]{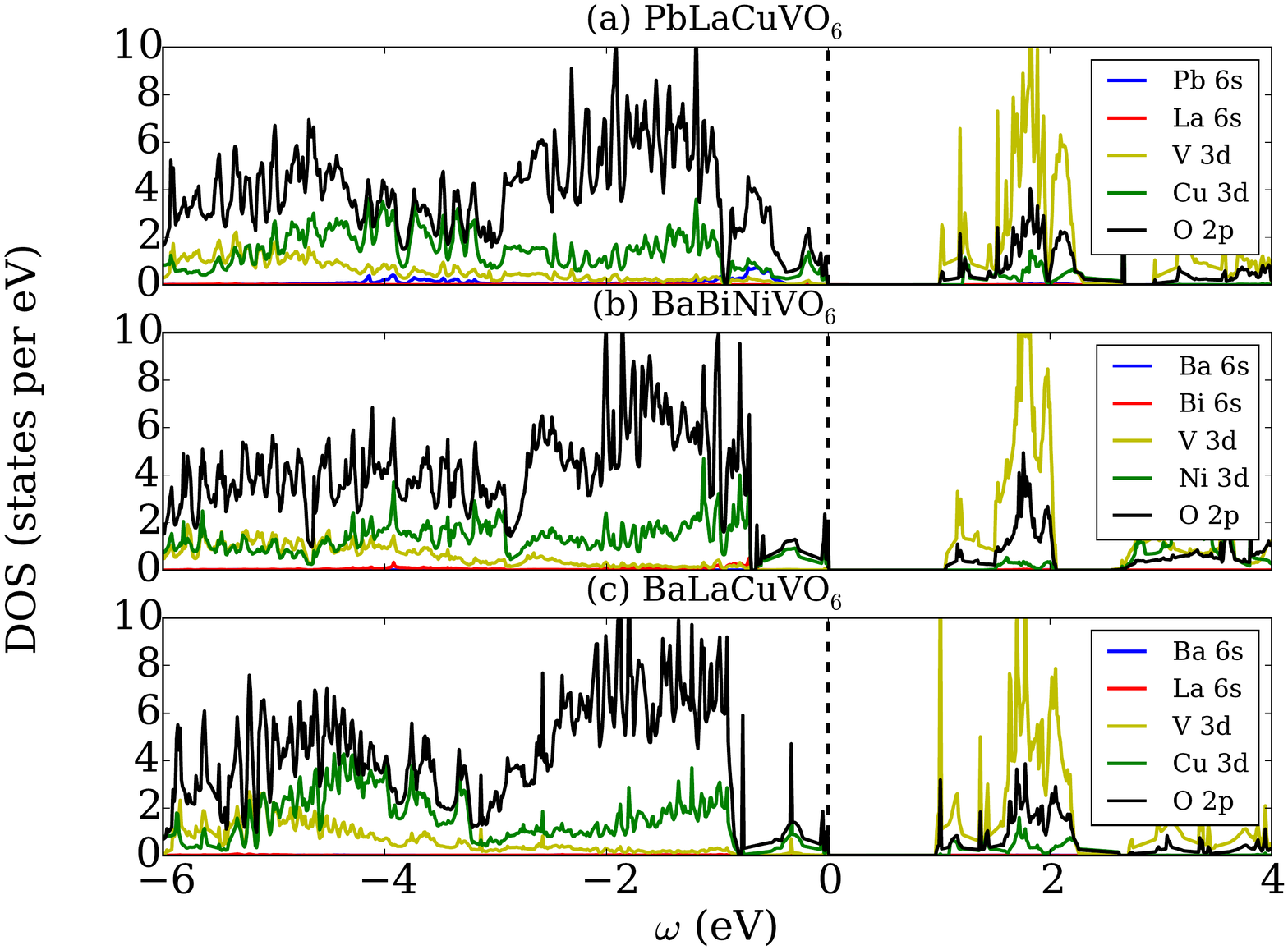}
\caption{{\bf $|$ Atomic resolved density of states of AA$^\prime$BB$^\prime$O$_6$} in the R$_{\vec q=\pi, \pi, 0}$ structure for {\bf (a)\hspace{1mm} } PbLaCuVO$_6$ {\bf (b)\hspace{1mm} } BaBiNiVO$_6$ {\bf (c)\hspace{1mm} } BaLaCuVO$_6$ ; DFT + U method is employed ( U$_{V}$ = 0, U$_{Cu}$ = 5 eV, U$_{Ni}$ = 5 eV); Fermi energy is set to be 0; }
\end{figure}